\documentclass[aps,showpacs,twocolumn,superscriptaddres]{revtex4-1}

\usepackage{graphicx}
\usepackage{bbm} 
\RequirePackage{amsmath} 
\usepackage{amssymb} 
\usepackage{color}
\usepackage[dvipsnames]{xcolor}
\usepackage[hidelinks]{hyperref}
\usepackage{natbib}

\usepackage{physics} 
\usepackage{comment}

\begin{document}

\title{ Geometric phase in a dissipative Jaynes-Cummings model:\\ theoretical explanation for  resonance robustness} 
\author{Ludmila Viotti}
\affiliation{Departamento de F\'\i sica {\it Juan Jos\'e
Giambiagi}, FCEyN UBA Facultad de Ciencias Exactas y Naturales,
Ciudad Universitaria, Pabell\' on I, 1428 Buenos Aires, Argentina.}
\affiliation{The Abdus Salam International Center for Theoretical Physics, Strada Costiera 11, 34151 Trieste, Italy}
\author{Fernando C. Lombardo}
\author{Paula I. Villar}
\affiliation{\it Instituto de F\'\i sica de Buenos Aires (IFIBA), CONICET-}
\affiliation {Departamento de F\'\i sica {\it Juan Jos\'e
Giambiagi}, FCEyN UBA Facultad de Ciencias Exactas y Naturales,
Ciudad Universitaria, Pabell\' on I, 1428 Buenos Aires, Argentina.}

\date{\today}

\begin{abstract}   
We follow a generalized kinematic approach to compute the geometric phases acquired in both unitary and dissipative Jaynes-Cummings models, which provide a fully quantum description for a two-level system interacting with a single mode of the (cavity) electromagnetic field, in a perfect or dissipative cavity respectively. In the dissipative model, the non-unitary effects arise from the outflow of photons through the cavity walls and the incoherent pumping of the two-level system.
Our approach allows to compare the geometric phases acquired in these models, leading to an exhaustive characterization of the corrections introduced by the presence of the environment. We also provide geometric interpretations for the observed behaviours. 
When the resonance condition is satisfied, we show the geometric phase is robust, exhibiting a vanishing correction under a non-unitary evolution. This fact is supported with a geometrical explanation as well.
\end{abstract}
\maketitle

\section{Introduction}

The existence of a geometric phase (GP) acquired by the state of a quantum system was discovered on theoretical grounds by Berry, in the context of adiabatic, cyclic, unitary evolution \cite{berry}. Thereafter it has been generalized to non-adiabatic cyclic, non-cyclic and even to non-unitary evolution \cite{aharonovanandan, samuel, pati, wilczek1984, anandan1988, tong2003, tong2004, Wu_2010}. All these generalizations reduce to the corresponding less-general results as more conditions are fulfilled. The GP was also shown to be a consequence of quantum kinematics and interpreted in terms of a parallel transport law that depends only on the geometry of the Hilbert space, from where it gets its name \cite{simon}.
As the extensive existing bibliography reflects, GPs have become not only a fruitful course of investigation to infer fundamental features of a quantum system but also of technological interest. 
For example, being robust to the fluctuations of a coupled bath, it was proposed as an important resource for the construction of phase gates \cite{jones, zanardi, schuster, lu, xiang2001nonadiabatic, zhu2002implementation, Sjoqvist_2012, xu2012nonadiabatic} in quantum information systems. 

A proper generalization for mixed state geometric phase under non-unitary evolution was presented in \cite{tong2004}. This definition has been used to measure the corrections induced on the GP in a  non-unitary evolution \cite{prl} and to explain the noise effects in the observation of the GP in a superconducting qubit \cite{leek,pra2014}. In particular, the GP of a two-level system under the influence of an external environment has been studied in a wide variety of scenarios \cite{042311,*707713,*VILLAR2009206,*052121,*032338}. It has further been used to track traces of quantum friction in both the very simplistic analytical model of an atom coupled to a scalar quantum field and the experimentally viable scheme of an atom traveling at constant velocity in front of a metallic surface \cite{epl,farias_nature, friccion2021}.

The GP accumulated by the state of an open quantum system will undoubtedly be different from that accumulated by the associated closed system, since the evolution is now plagued by non-unitary effects such as decoherence and dissipation. It is commonly said that the coupling of the quantum system corrects the unitary GP, by noting that $\phi_g = \phi_g^u + \delta\phi$, being $\delta\phi$ proportional to the coupling of the system and the environment. Under suitable conditions, these corrections can be measured by means of an interferometric (atomic interference) \cite{ivette,moore}, spin echo \cite{leek}, and NRM \cite{Du, prl} experiments. 

The quantum Rabi model, which represents the dipole interaction between a two-level atom and the electromagnetic (cavity) field, is a paradigmatic model that has been widely used in many areas of research, ranging these from quantum optics and quantum information science to condensed matter physics. After performing the Rotating Wave Approximation (RWA) it is possible to obtain the Jaynes-Cummings (JC) model, which was first used in 1963 to examine aspects of spontaneous emission and to reveal the existence of Rabi oscillations in atomic excitation probabilities, for fields with sharply defined number of photons. It may be considered the simplest elementary model that successfully accounts for the interaction of radiation with matter. Despite being extremely simple and analytically solvable, it manages to explain many of the cavity electrodynamics experiments to date, as well as more recent experiments on superconducting qubits. Moreover, further generalizations in both coherent and dissipative models have been investigated lately from different approaches. \cite{MSA, dong2015, zhu2016}

Geometric phases on JC and quantum Rabi models have been investigated to a considerable extent both theoretically  \cite{ivette,carollo_5,larson,jia} and experimentally \cite{pechal}. Much of the work has been done in the context of adiabatic evolution, computing the vacuum-induced Berry phase acquired by the instantaneous eigenstates of a spin-1/2 particle which interacts with an external magnetic field, while the  direction of the field is slowly changed in a cyclic fashion. In \cite{ivette} authors  generalized that original model to its full quantum counterpart, wherein the classic driving field was replaced by a quantized field, without abandoning neither the cyclic nor the adiabatic conditions.

In the present work we shall make a thorough study of the GP acquired in a dissipative JC model, where the two-level system (TLS) is interacting with a single mode of the quantized electromagnetic field, in a dissipative cavity. We consider the interaction between the atom-mode system and its environment to be given by the flow of photons through the cavity mirrors and the continuous and incoherent pumping of the TLS. This is a frequent scenario in semiconductor cavity quantum electrodynamics (QED) \cite{kira}.
In order to be able to compare the GPs emerging on this model with those accumulated in unitary evolution, we need to set an approach that allows for the computation of GPs in the case time evolution due to JC Hamiltonian and that has a direct generalization to non-unitary evolution. 
It is Mukunda and Simon's kinematic approach \cite{mukunda93} for the computation of GPs when the system is subject to unitary, but otherwise general evolution that enables us to address the matter. The program is completed by recurring to the already mentioned consistent generalization of the GP proposed by Tong et. al. \cite{tong2004}.

The article is organized as follows. In section \ref{sec:unit} we show in detail the calculation of the GP in the unitary case. For this purpose, in Sec. \ref{sec:bp} we review the usual approach to the subject, computing the Berry phase accumulated when the system is driven by a phase-shift operator. As for computing the GP acquired in temporal evolution, we later introduce a different generalized approach, the kinematic approach proposed by Mukunda and Simon \cite{mukunda93}. This generalized geometric phase will be present and applied in Sec \ref{sec:up}.
In Sec. \ref{sec:nonunit} the dissipative model for the JC system is presented, in conjunction with the master equation governing its dynamics. We shall numerically study the evolution of the system and the decoherence process evidenced in the decay in the coherences of the reduced density matrix for different conditions given by different parameters values.  In this section, the GP in the 
non-unitary regime is studied and compared with that obtained for unitary evolution. Finally, in Sec. \ref{sec:conclu} we summarize our conclusions.


\section{Unitary Jaynes-Cummings model}\label{sec:unit}
The quantum Rabi model, which represents the simplest interaction between a two-level atom and the electromagnetic (cavity) field, is a paradigmatic model that has been widely used in many areas of research, ranging these from quantum optics and quantum information science to condensed matter physics. The Hamiltonian is given by
\begin{equation}
\hat{H}_{\rm R} = \omega \hat{a}^\dagger \hat{a} + \frac{\epsilon}{2}\sigma_z + \lambda ( \hat{a}^\dagger + \hat{a} )\sigma_x, \label{rabi}
\end{equation}
where $\epsilon$ and $\omega$ are the natural frequencies of two-level system and the cavity mode respectively, $\sigma_{x,y,z}$ are the usual Pauli matrices acting on the two atomic internal states $\ket{g}$ and $\ket{e}$, $\hat{a}$ ($\hat{a}^\dagger$) are the annihilation (creation) photon operators, and $\lambda$ is the effective coupling strength between the system and the field, taken to be real. Therefore, near resonance ($\epsilon \approx \omega$), and if $\lambda$ is much smaller than the natural frequencies of the system, the RWA is justified and the interaction Hamiltonian is now given by the so called JC Hamiltonian 
\begin{equation}
\hat{H}_{\rm JC} = \omega \hat{a}^\dagger \hat{a} + \frac{\epsilon}{2}\sigma_z + \lambda ( \hat{a}^\dagger \sigma_-+ \hat{a} \sigma_+),
\label{jc}\end{equation}
where $\sigma_\pm = (\hat{\sigma}_x \pm \mathrm{i} \hat{\sigma}_y)/2$ are the atomic raising and lowering operators. 

It is usual to perform the following unitary transformation $\hat{K}$ on $\hat{H}_{\rm JC}$ \cite{luo} 

\begin{equation} 
\hat{K} = \exp\left[-\mathrm{i} \omega t (\hat{a}^\dagger \hat{a} + \frac{\sigma_z}{2})\right].
\end{equation}
Thus, the new JC Hamiltonian is then given by, 

\begin{equation}
\hat{H} =  \frac{\Delta}{2}\sigma_z + \lambda ( \hat{a}^\dagger \sigma_-+ \hat{a} \sigma_+),\label{jcsinw}
\end{equation}
with $\Delta = \epsilon - \omega$ denoting the atom-field detuning. Resonance condition implies $\Delta = 0$. 
Herein, we shall restrict the calculation to the subspace generated by the base ${\cal B} = \{\ket{e, n}, \ket{g, n+1} \}$. In this case,  the eigenvalues and eigenstates of the Hamiltonian of Eq.(\ref{jcsinw}) are given, respectively, by 
\begin{equation} 
E_{\pm}^{(n)} = \pm \frac{1}{2}\sqrt{ 4 \lambda^2 (n + 1) + \Delta^2 },
\end{equation}
and 
\begin{eqnarray} 
\ket{\Phi_n^-} &=& \sin\frac{\theta_n}{2} \ket{e, n} - \cos\frac{\theta_n}{2}  \ket{g, n+1} \nonumber \\
\ket{\Phi_n^+} &=& \cos\frac{\theta_n}{2} \ket{e, n} + \sin\frac{\theta_n}{2}  \ket{g, n+1} ,
\label{eq:eigen}
\end{eqnarray}
where $\cos\theta_n = \Delta/\sqrt{4\lambda^2 (n+1) + \Delta^2}$. 
It is important to mention that if resonance condition $\Delta =0$ is satisfied, the eigenenergies of the system reduce to $E_{\pm}^{(n), \Delta=0}=\pm \sqrt{n + 1}\;\lambda$ and the associated eigenstates correspond to Bell states
\begin{equation} 
   \ket{\Phi_n^\pm} = \frac{1}{\sqrt{2}} (\ket{e, n} \pm \ket{g, n+1}).
    \label{eq:Bell}
\end{equation}
We shall return to this issue later in the manuscript.


\subsection{Berry phase}\label{sec:bp}
The seminal work of Fuentes-Guridi et.al. \cite{ivette} and a number of later works \cite{luo, pechal, wang2015} show that, in the isolated (unitary) case, the eigenstates of the system acquire a nontrivial Berry phase when driven by an adiabatic phase transformation, even in the vacuum field-state. Performing an unitary transformation $\hat{R}(\varphi) = \exp[-i \varphi \hat{a}^\dagger \hat{a}]$ on the Hamiltonian $\hat{H}$ we get $\hat{H}(\varphi)$
\begin{equation}
\hat{H}(\varphi) =  \frac{\Delta}{2}\sigma_z + \lambda ( \hat{a}^\dagger e^{-\mathrm{i} \varphi} \sigma_-+ \hat{a}  e^{\mathrm{i} \varphi}\sigma_+),\label{jcconphi}
\end{equation}
where $\hat{R}(\varphi)$ is a phase-shift operator of the field and $\varphi$ is an external control parameter. The eigenstates $\ket{\Phi_n^\pm(\varphi)}$ of $\hat{H}(\varphi)$ can be obtained as $\ket{\Phi_n^\pm(\varphi)} = \hat{R}(\varphi)\ket{\Phi_n^\pm}$ with $\ket{\Phi_n^\pm}$ being the eigenstates of $\hat{H}$. Varying $\varphi$ slowly from $0$ to $2\pi$, 
\begin{eqnarray}  \nonumber 
\phi_g^u &=& \mathrm{i} \int_{\cal C} d\varphi \, \bra{\Phi_n^\pm(\varphi)} \frac{d}{d\varphi} \ket{\Phi_n^\pm(\varphi)} \label{unitaria} , (n = 0, 1, 2, ...)\\
&= & \pi \left[ 1 \mp \frac{\Delta}{\sqrt{4\lambda^2 (n+1) + \Delta^2}}\right] + 2 n \pi,
\label{eq:bp}
\end{eqnarray}
where the phase is non-trivial for $n=0$, which means that even the vacuum state of the field introduces a correction in Berry’s phase. As for the dependence on the atom-mode detuning, it is worth noting that in the resonant ($\Delta=0$) case, the evolution is restricted to states of the form
\begin{equation} 
   \ket{\Phi_n^\pm} = \frac{1}{\sqrt{2}} (\ket{e, n} \pm e^{-\mathrm{i} \varphi}\ket{g, n+1}),
    \label{eq:Bell_Berry}
\end{equation}
corresponding to maximally entangled states of a pair of TLSs. In this case, it has already been demostrated that the bipartite system acquires a well-stated $\pi$ phase \cite{Milman,Oxman_fractional,erikreview}.

\subsection{Geometric phase accumulated in time and the kinematic approach}\label{sec:up}
With the future purpose of contrasting the unitary and dissipative dynamics of the system by inspecting the GPs acquired in both cases, we turn now to a more general approach to geometric phases than that of Berry's. The main reason for pursuing this is rooted in the need of studying the phases aquired during the temporal evolution governed by $\hat{H}_{\text{JC}}$, which particularly result in no GPs accumulated on the states $\ket{\Phi_n^\pm}$  when  Berry's approach is followed. 

We address the study of GPs emerging from the evolution generated by $\hat{H}_{\text{JC}}$ by adopting Mukunda and Simon's kinematic approach \cite{mukunda93}. In this approach the GP associated with a general open curve in the space of states is defined in a gauge and reparametrization invariant way and therefore derived as a natural consequence of the quantum kinematics. We begin by briefly introducing the formalism and, afterwards, turn to its application on the system under examination.

\begin{figure}[htbp]
\centering
\includegraphics[width=.8\linewidth]{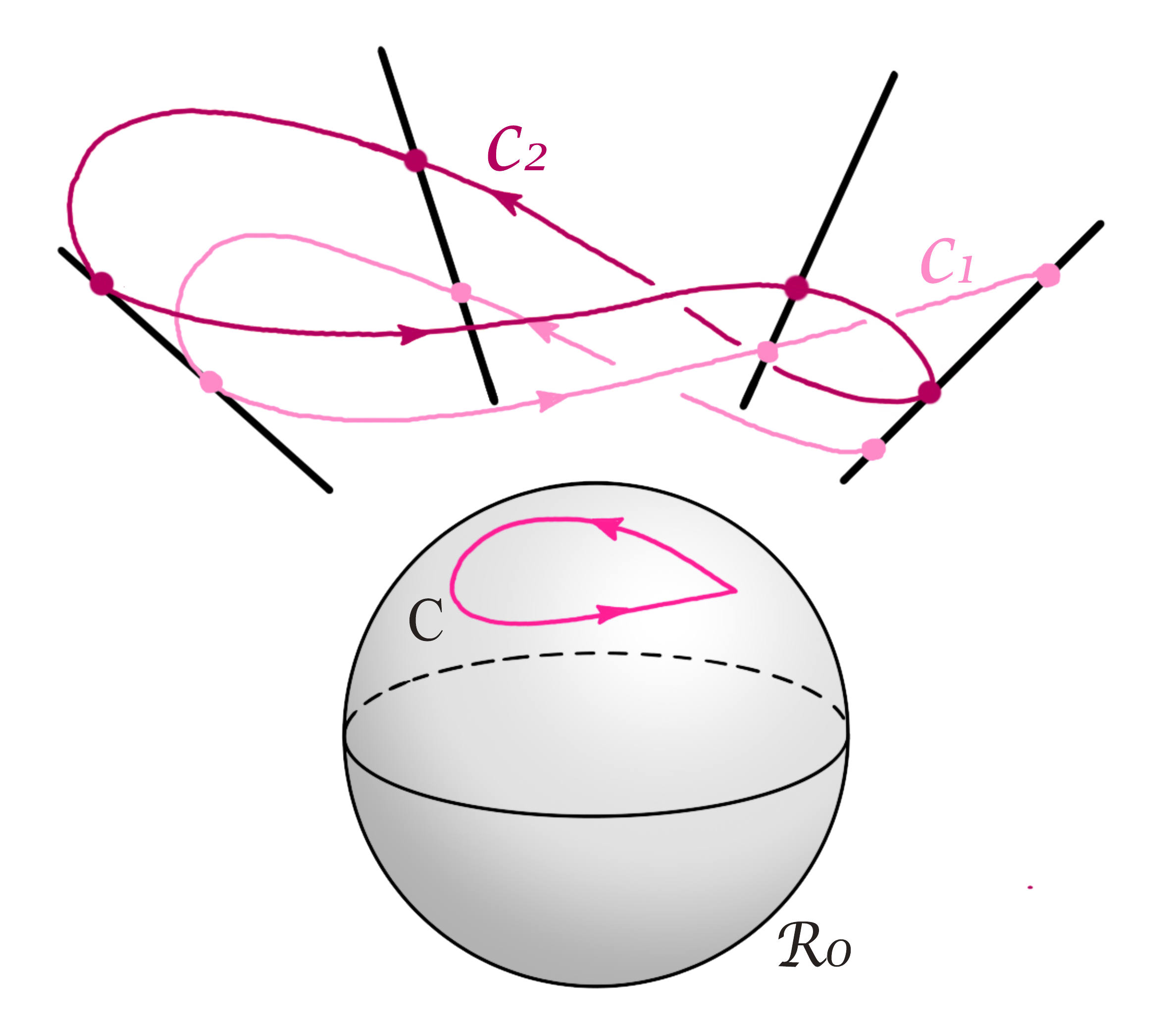}
\caption{Schematic representation of the ray space $\mathcal{R}_0= \mathcal{H}_0/U(1)$ composed of classes of equivalent Hilbert space vectors. A curve C described by the physical state of the system on $\mathcal{R}_0$ can thus have many different liftings into the space of unitary vectors $\mathcal{H}_0$.}
\label{fig:Hilbert}
\end{figure}

Let $\mathcal{H}$ be a Hilbert space suitable for the description of some quantum system and $\mathcal{H}_0$ the subspace of $\mathcal{H}$ containing all the unit vectors. The equivalence relation $\ket{\psi}\sim\ket{\psi'}=e^{\mathrm{i}\alpha}\ket{\psi}$ between elements $\{\ket{\psi}\in\mathcal{H}_0\}$ favors the definition of the ray space $\mathcal{R}_0= \mathcal{H}_0/U(1)$, where each element of $\mathcal{R}_0$ represents a physically distinct state. For a two-level system, the ray space is known as Bloch sphere and is depicted in Fig. \ref{fig:Hilbert}.
As it evolves, the state of the system may describe a curve $\text{C}$ on the ray space, which can be lifted to a curve $\mathcal{C}=\{\ket{\psi(t)}\in \mathcal{H}_0\;|\; t\in [t_1,t_2] \subset \mathrm{R}\}$ in the Hilbert space. Since the relation between the elements of $\mathcal{H}_0$ and $\mathcal{R}_0$ is not one-to-one, there will be many possible liftings $\mathcal{C}$ corresponding to the same path C. Fig. \ref{fig:Hilbert} illustrates this multiplicity of liftings as well.

The geometric phase associated with the curve $\text{C}$ in the ray space can be written in terms of some lift into vectors of the Hilbert space as

\begin{equation}
    \phi_g[\text{C}] = \arg \left( \bra{\psi(0)} \ket{\psi(t)} \right) - {\rm Im} \int_0^t dt' \bra{\psi(t')} \ket{\dot{\psi}(t')} \label{fase:u}
\end{equation} 
where each term on the r.h.s. of Eq. (\ref{fase:u}) depends on the particular set of Hilbert space vectors $\{\ket{\psi(t)}\}$ of the lift, while the combination depends solely on $\text{C}$ \cite{acotacion}. 
It is important to remark that, in the generalization described, each previous result is recovered if stronger conditions are satisfied. For example, it is easy to demonstrate that for cyclic evolution $\phi_g[\text{C}]$ reduces to the Aharonov-Anandan formula \cite{aharonovanandan}, and to the Berry phase for adiabatic and cyclic evolution.

As already mentioned, if the system is prepared in an eigenstate $\ket{\Phi_n^\pm}$ and evolved according to $\hat{H}$, its state at later times $e^{\pm \,i\,E_+\,t}\ket{\Phi_n^\pm}$ will be represented by a dot in the Bloch sphere and will acquire no geometric phase.  
But as any path in the ray space can be examined, the computation of GPs acquired by any state is allowed. In particular, it is possible to consider states which are not instantaneous eigenstates of $\hat{H}_{\text{JC}}$.
On the other hand, if we consider the initial state to be $\ket{\psi(0)} = \ket{e, n}$, it can be seen that the temporal evolution leads to a state 

\begin{align}\nonumber
    \ket{\psi(t)} = \left(\cos^2(\theta_n)e^{-i E_+\,t} + \sin^2(\theta_n) e^{i \, E_+\, t}\right)\ket{e,n}\\[.75 em]
    - i\,\sin(\theta_n)\sin(E_+ \,t)\ket{g, n+1},
\end{align}
which describes a curve on the Bloch sphere, accumulating a phase

\begin{align}\nonumber
    \phi_g(t)= &\pi\cos(\theta_n)\frac{t}{\tau}
    -\pi\left[\frac{t}{\tau}+\frac{1}{2}\right]\\[0.75 em]
    &-\arctan\left\lbrace\cos(\theta_n) \tan\left(\pi\frac{t}{\tau}\right)\right\rbrace.
\end{align}
Here $\tau =2\pi/\Omega_n$ is the period defined by the Rabi frequency $\Omega_n= \sqrt{\Delta^2 + 4\lambda^2(n+1)}$ of the atom-mode system, and hard brackets $[\cdot]$ denote integer part. It is worth seeing that, when $t=\tau$ the geometric phase accumulated is

\begin{equation}
    \phi_g(\tau)=-\pi\left(1 - \frac{\Delta}{\sqrt{\Delta^2 + 4\lambda^2(n+1)}}\right).
\end{equation}

\begin{figure}[htbp]
\centering
\includegraphics[width=.8\linewidth]{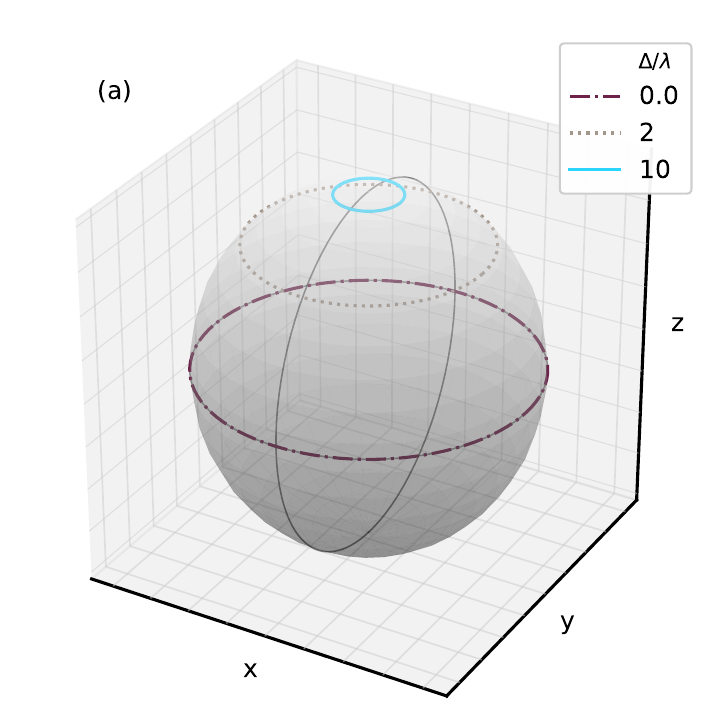}
\includegraphics[width=.8\linewidth]{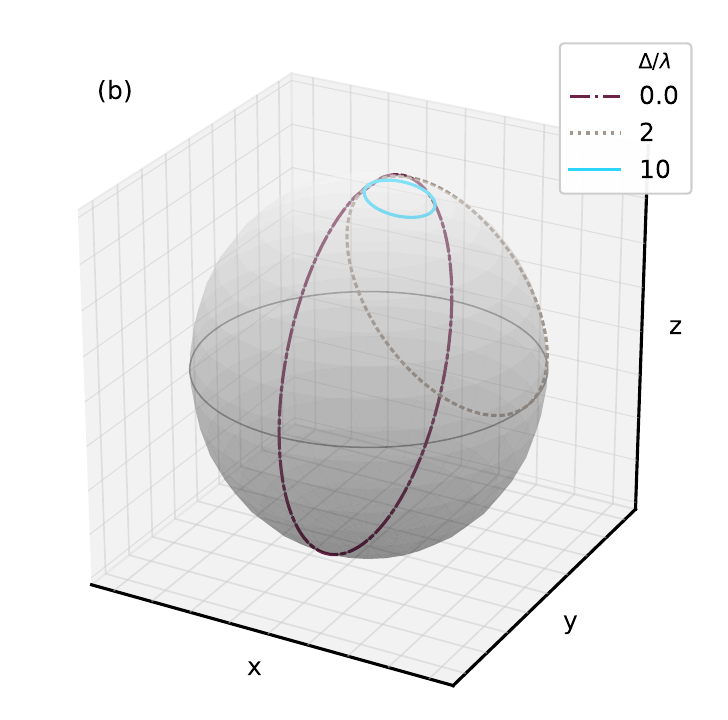}
\caption{Trajectories described by the state of the system for different values of $\Delta/\lambda$: (a) corresponds to evolutions of the eigenstate $\ket{\Phi_0^+}$ under the action of a phase-shift operator, while (b) corresponds to evolutions of $\ket{e,0}$ according to $H_{\text{JC}}$.}
\label{fig:BlochUnit}
\end{figure}

The coincidence up to a minus sign with the usual result expressed in Eq. (\ref{eq:bp}) can be explained by comparing the curves described in the Bloch sphere for each evolution, which are displayed in Fig. \ref{fig:BlochUnit}. Figure \ref{fig:BlochUnit}.a shows the path traced by the state $\ket{\Phi_0^+(\varphi)}$ of a system which is prepared in the eigenstate $\ket{\psi(0)}=\ket{\Phi_0^+}$ and evolves under the action of a phase-shift operator of the field $\hat{R}(\varphi)$. In this well-known situation the state describes latitude circles, corresponding to the equator in the resonant case.
Fig. \ref{fig:BlochUnit}.b shows the curves described by the state $\ket{\psi(t)}$ of a system which is initially prepared in the state $\ket{\psi(0)}=\ket{e,0_c}$ and evolves according to $H_{\text{JC}}$. Under these circumstances, the state of the system describes circular arcs containing the north pole, which represents the state $\ket{e,0_c}$. It can be seen that, for each value of $\Delta/\lambda$ the curves described in each evolution are related by a rigid rotation followed by a reflection. Rotations of the Bloch sphere correspond to isometries that can be realized through an unitary operator acting on the Hilbert space. As this rotation applies to the whole curve C, it follows that it can be described with an unitary static transformation that leaves the GP invariant. On the other hand, reflections are isometries that can be realized through anti-unitary lifts into the Hilbert space and reverses the GP \cite{mukunda93,samuel2}.
In addition, if we particularly focus on the resonant case, it can be seen from the above explanation that the temporal evolution of $\ket{\psi(t)}$ is restricted to a subspace of $\mathcal{H}_0$ related to that of Eq. (\ref{eq:Bell_Berry}) through the mentioned unitary static operator, and thus it is likely to acquire the same phase (up to a minus sign).

\section{Non-unitary evolution of the system}\label{sec:nonunit}

In cavity QED the main source of dissipation is the leakage of photons through the cavity mirrored walls. A second source of dissipation and decoherence is the pumping of the TLS, a common situation which is usually neglected.
We will address the problem of the description of the dissipation and decoherence processes in a two-level atom-cavity system at zero temperature considering both dissipative mechanisms through a phenomenological master equation \cite{MSA}. In Lindblad form, the equation is
\begin{equation}
\dot\rho = -i [\hat{H} , \rho(t)] + \gamma {\cal D}[a, a^\dagger] \rho(t) + p {\cal D}[\sigma_\pm] \rho(t), 
\end{equation}
where $\gamma$ is the rate of leakage of photons out of the cavity and $p$ is the amplitude of the continuous pump. 
The explicit phenomenological equation for $\rho(t)$, given by

\begin{eqnarray}
\dot\rho &=& -i [\hat{H} , \rho] + \frac{\gamma}{2} \left[ 2 a \rho a^\dagger - a^\dagger a \rho - \rho a^\dagger a\right] \nonumber \\
&+&  \frac{p}{2} \left[ 2 \sigma_+\rho \sigma_- - \sigma_- \sigma_+ \rho - \rho  \sigma_- \sigma_+\right],
\label{eq:master}
\end{eqnarray}
will be studied in the case of low pumping ($p \ll \lambda, \gamma$).

According to the relationship between $\lambda$ and $\gamma$ there are two clearly differentiated regimes \cite{carmichael1989, review_lodahl2015, book_yamamoto2003, vera2009characterization, laussy2008strong}. The first one, known as the strong coupling (SC) regime, is characterized by the fact that the interaction constant is bigger than the system dissipation rate ($\lambda > \gamma$). In the second one, known as the weak coupling (WC) regime, the opposite occurs, i.e., the interaction constant is smaller than the dissipation rate of the system ($\lambda < \gamma$). We stress that in an unitary theory considering a bipartite system, the most natural definition of weak and strong coupling would depend on the ratio between the constant characterizing the coupling between sub-systems ($\lambda$ for us) and those magnitudes characterizing the internal dynamics of the individual systems. In the present work, however, we are considering this composite system is subjected to dissipative effects, arising from the interaction of both individual subsystems with the environment. When studying this models it is quite common to name the emerging regimes as we have done.
For the description of any of these regimes, we assume that the atom can be in its ground $|g\rangle$ or excited $|e\rangle$ state, and the cavity photonic field have zero $| 0_c\rangle$ or one photon  $| 1_c\rangle$. Therefore, the atom-cavity system is described by the bare states, 
$| 0 \rangle = | g, 0_c\rangle$, $| 1 \rangle =  | e, 0_c\rangle$, and $| 2 \rangle =  | g, 1_c\rangle$. Explicitly computing each element of Eq. (\ref{eq:master}), the set of equations describing the dynamics is,

\begin{eqnarray}
\dot{\rho}_{00} &=& - p \rho_{00} + \gamma \rho_{22} \nonumber \\
\dot{\rho}_{11} &=& - i \lambda \left(\rho_{21} - \rho_{12}\right) + p \rho_{00} \nonumber \\
\dot{\rho}_{22} &=&   - i \lambda \left(\rho_{12} - \rho_{21}\right)  - \gamma  \rho_{22} \nonumber \\
\dot{\rho}_{12} &=&   - i \lambda \left(\rho_{22} - \rho_{11}\right) - i \Delta \rho_{12} - \frac{\gamma}{2} \rho_{12} \nonumber\\
\dot{\rho}_{01} &=&   - \frac{p}{2} \rho_{01} + i (\Delta \rho_{01} + \lambda \rho_{02}) \nonumber\\
\dot{\rho}_{02} &=&   i \lambda \rho_{01} - \frac{1}{2}(p+\gamma) \rho_{02}, \label{master}
\end{eqnarray}
where $\rho_{ji} = \rho^*_{ij}$.

\begin{figure}[ht]
\centering
\includegraphics[width=.9\linewidth, trim={.75cm 0 .75cm 0}]{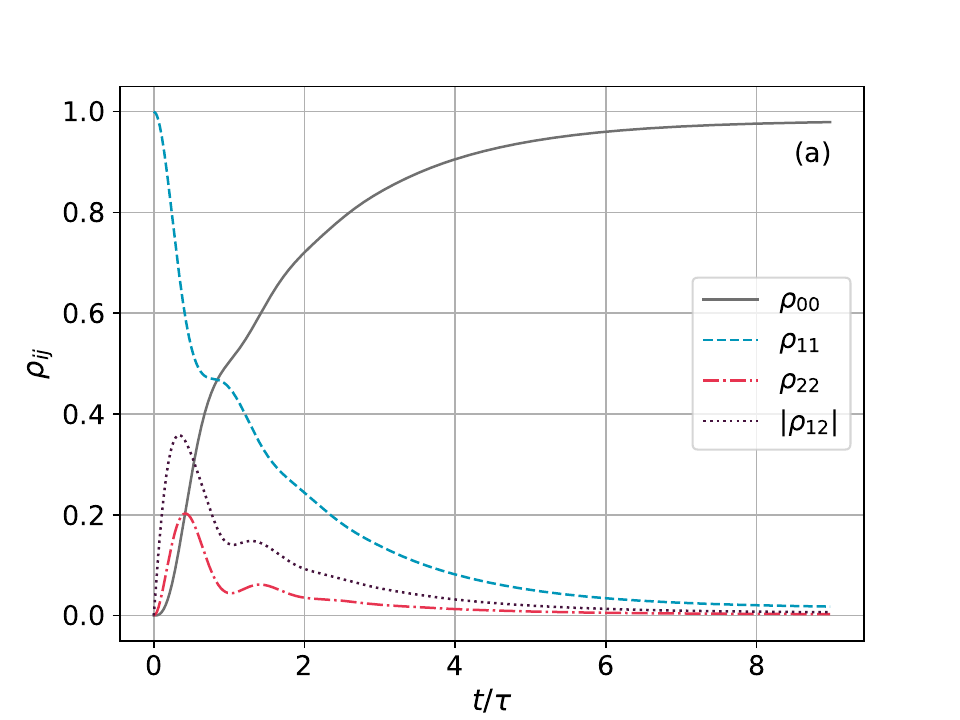}
\includegraphics[width=.9\linewidth, trim={.75cm 0 .75cm 0}]{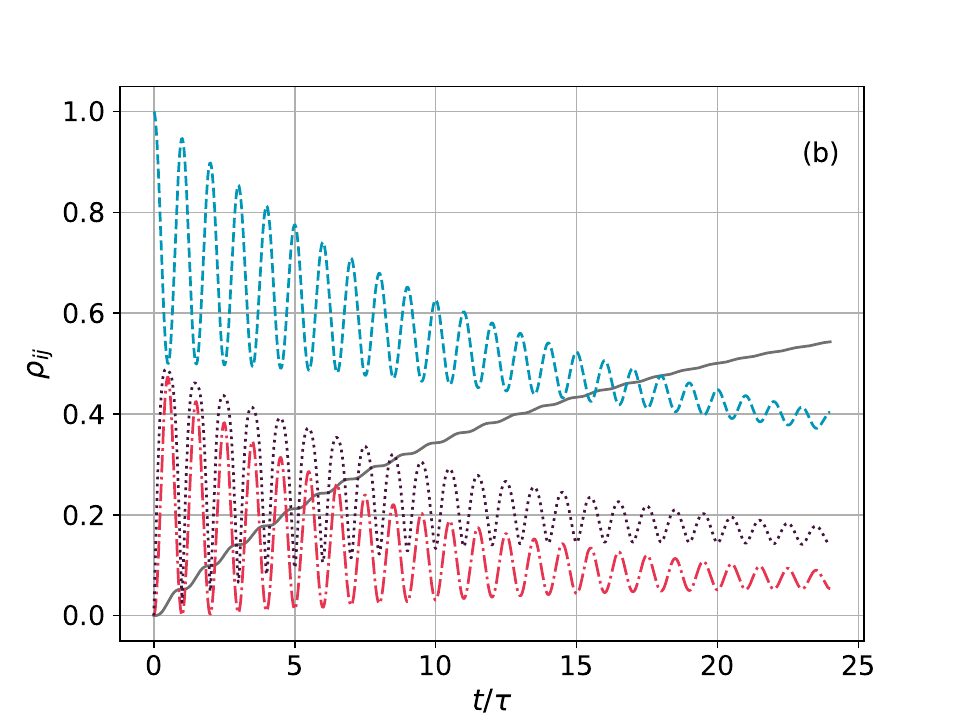}
\caption{Dynamics exhibited by the matrix elements $\rho_{ij}$ in the (a) weak coupling regime characterized by $\gamma/\lambda = 2$ and (b) strong coupling regime with $\gamma/\lambda = 0.1$. Solid, dashed and dot-dashed lines correspond to populations $\rho_{00}$, $\rho_{11}$ and $\rho_{22}$ respectively , while the dotted lines show the imaginary part of the coherence element $\rho_{12}$ in each case. In the WC regime (a), the system loses coherence and quickly tends to an asymptotic state $\rho \sim \ket{0}\bra{0}$ of minimal excitation. On the other hand, in the SC regime (b) the state preserves coherence for many pseudo-cycles and evolves to an asymptotic mixed state. Parameter values are  $\Delta/\lambda = 2$, $p/\lambda= 0.005$.}
\label{fig:elementos}
\end{figure}

As the elements $\rho_{01}$ and $\rho_{02}$ are decoupled from the rest, any solution $\rho(t)$ of Eq. (\ref{master}) with initial conditions implying $\rho_{0i}(0)=0$ will have $\rho_{00}(t)$ as the only non-zero element of its 0-file and 0-column. Restricting to this kind of initial conditions, we may write

\begin{equation} {\rho(t)} = \begin{pmatrix}
\rho_{00} & \begin{array}{cc}
                 0 & 0 
            \end{array}\\
\begin{array}{c}
     0\\
     0 
\end{array} & \tilde{\rho}_{2\times2}
\end{pmatrix},
\label{eq:block}
\end{equation}
where 
\begin{equation}
{\tilde \rho} = \begin{pmatrix}
\rho_{11}& \rho_{12} \\
\rho_{21} & \rho_{22} 
\end{pmatrix}.
\end{equation}
By  solving  numerically the remaining equations, we can find the state $\rho(t)$ at later times. Fig. \ref{fig:elementos} shows the evolution of the density matrix elements $\rho_{ij}(t)$ for a system which is initially prepared in the state $\rho(0)=\ket{e,0_c}\bra{e,0_c}$. It can be seen that in the WC regime ($\gamma > \lambda$), implying that the coupling to the environment is stronger than the coupling responsible of transitions in the $\{\ket{g, 0_c}, \ket{e, 0_c}, \ket{g, 1_c}\}$ space, the system loses coherence and quickly tends to an asymptotic state $\rho \sim \ket{0}\bra{0}$ of minimal excitation. On the other hand, in the SC regime ($\gamma < \lambda$), the state preserves coherence for many pseudo-cycles of length $\tau = 2\pi/\Omega_0$ and evolves to an asymptotic mixed state.

It is important to mention that the observed behaviour of the state $\rho(t)$ on each regime, determines that the SC is more suitable for the study of the GP. Thus, in the following, WC regime will be necessarily disregarded.

\subsection{Non-unitary case: the vacuum-induced corrections to the geometric phase}\label{sec:gpnonunit}

In the following, we shall concentrate on how the GP acquired is corrected when the bipartite system state evolves in presence of noise and dissipation. 
For a mixed state under a non-unitary evolution the GP acquired is defined as \cite{tong2004}

\begin{eqnarray} 
\phi_g(t) &=& {\rm Arg}\left\{\sum_k \sqrt{ \varepsilon_k (0) \varepsilon_k (t)}
\bra{\Psi_k(0)}\ket{\Psi_k(t)} \right. \nonumber \\
&\times & \left.  \exp{\left[-\int_0^{t} dt' \bra{\Psi_k(t')}
\ket{\Dot{\Psi}_k(t')} \right]}\right\}, \label{fase:nonu}
\end{eqnarray}
where $\varepsilon_k(t)$ and $|\Psi_k\rangle$ are the instantaneous eigenvalues and eigenstates of the reduced density matrix $\rho$ that is solution of the master equation. 
It is worth noting that the phase in Eq.(\ref{fase:nonu}), even though defined for non degenerate mixed states, reduces to that given by Eq. (\ref{fase:u}) for pure states under unitary evolution. 
Thanks to the block diagonal form Eq.(\ref{eq:block}) displayed by the examined density matrices, we only have to consider the $2\times2$ block $\tilde{\rho}(t)$ in order to diagonalize $\rho(t)$.
Therefore, the eigenvalues of the total reduced density matrix $\rho(t)$ are
\begin{eqnarray}
\epsilon_0 &=& \rho_{00}(t) \\
\epsilon_\pm &=& \frac{1}{2} \left(\rho_{11} + \rho_{22} \pm
   \sqrt{(\rho_{11} - \rho_{22})^2 + 4 \rho_{12} \rho_{21}}\right). \nonumber 
   \end{eqnarray}

When the initial state of the system is pure, $\epsilon_k(0) = 1$ and $\epsilon_j(0) = 0 \, \forall j\neq k$, so the geometric phase defined in Eq.(\ref{fase:nonu}) reduces to 

\begin{equation} 
\phi_g(t) = {\rm arg}\left\{\bra{\Psi_+(0)}\ket{\Psi_+(t)} \right\} - {\rm Im} \int_0^{t} dt' \bra{\Psi_+(t')}
\ket{\Dot{\Psi}_+(t')}
\label{fase:iniu}
\end{equation}
which resembles Eq. (\ref{fase:u}). In fact, Eqs. (\ref{fase:u}) and (\ref{fase:iniu}) coincide except for the replacement of $\ket{\psi(t)}$ by the eigenstate $\ket{\Psi_+(t)}$. Thus, under the condition of a pure initial state, the phase accumulated admits the interpretation of being the unitary GP acquired by the eigenvector $\ket{\Psi_+(t)}$ of the system state $\rho$.

\begin{figure}[ht]
\centering
\includegraphics[width=.9\linewidth, trim={.75cm 0 .75cm 0}]{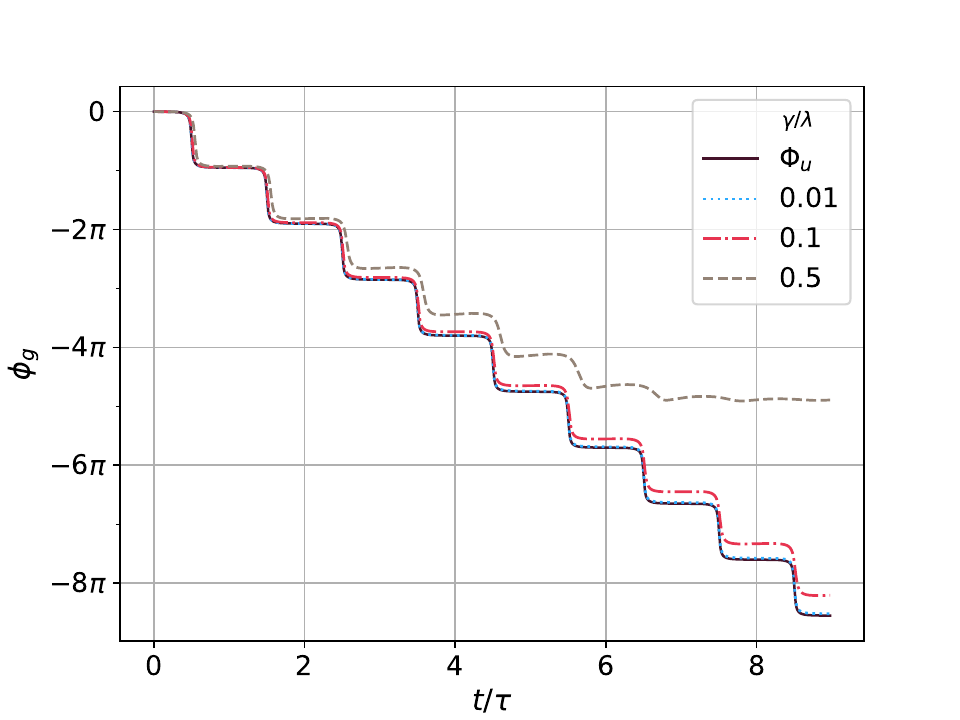}
\caption{Geometric phase $\phi_g (t)$ accumulated over time for different $\gamma/\lambda$ values, corresponding to different  rates of leakage of photons out of the cavity. Parameter values are $\Delta/\lambda = 0.1$, $p/\lambda= 0.005$ for every curve but the unitary case (solid line) for which $\Delta/\lambda = 0.1, p/\lambda=\gamma/\lambda= 0$.}
\label{fig:FvsTGamma}
\end{figure}

Let us consider, for example, an initial condition such that $\rho(0) = \ket{e, 0_c} \bra{e, 0_c}$, under which the geometric phase can be written as 
\begin{equation}
\phi_g(t) = \int_0^{t} dt' \, \frac{{\rm Im} \left(\dot{\rho}_{12}^*\,\rho_{12}\right)}{(\rho_{22} - \varepsilon_+)^2 + \rho_{12}\rho_{21}}. 
\label{fase:final}
\end{equation}
In general, this phase will differ from that acquired in unitary evolution since the leakage of photons and the continuous pumping of the atom introduce non-unitary effects such as decoherence and dissipation. This means the phase can be written as $\phi_g = \phi_u + \delta\phi$ where $\delta\phi$ is the correction to the unitary phase induced by the presence of the environment.
The difference $\delta\phi(t)$ between the the GP accumulated over time in the dissipative case and that accumulated in unitary evolution grows with $\gamma/\lambda$, as it can be observed in Fig. \ref{fig:FvsTGamma}, where the temporal evolution of the GP is displayed for different $\gamma/\lambda$ values. 

\begin{figure}[ht]
\centering
\includegraphics[width=.85\linewidth, trim={.5cm 0 .5cm 0}]{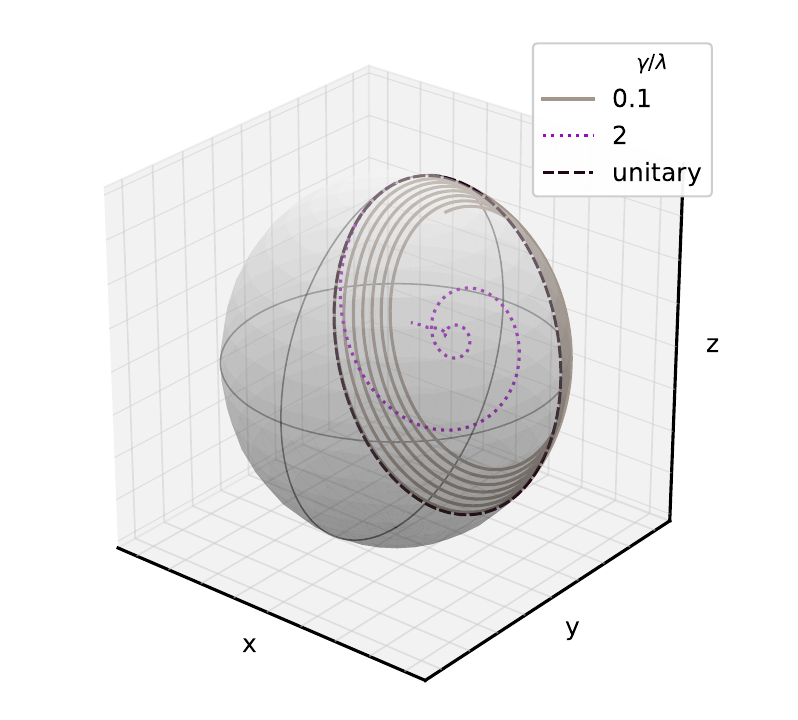}
\caption{Trajectories described on the Bloch sphere by the unitarily evolved state $\ket{\psi(t)}$ and the eigenstate $\ket{\Psi_+(t)}$ of $\rho(t)$ during $t= 7\tau$ evolutions. The unitary curve and those described in both regimes of the dissipative case differ, and thus the GPs accumulated over time differ as well. However, while in SC regime the motion last for the whole time interval, in WC regime the state stops any motion almost immediately.  Parameter values are $\Delta/\lambda = 1, p/\lambda= 0.005$ for non-unitary curves.}
\label{fig:BlochEigen}
\end{figure}

However, for large enough values of $\gamma/\lambda$, the loss of coherence stops the motion of the state in the ray space, and therefore the accumulation of GP is also halted. The extreme situation occurs in the WC regime, in which the state looses all coherence before accumulating almost any phase. The immediate loss of coherence in the WC regime was already evident in Fig. \ref{fig:elementos} by means of the matrix elements. That dynamics of the elements $\rho_{ij}$ manifest itself in the GP through the evolution of the eigenstate $\ket{\Psi_+(t)}$, as the GP acquired depends only on the path described by $\ket{\Psi_+(t)}$ on the ray space. A pictorial representation of this path can be found in Fig. \ref{fig:BlochEigen}, which shows the curve described by $\ket{\Psi_+(t)}$ on the Bloch sphere during three different $t= 7\tau$ evolutions, allowing for comparison of the three cases. The cases displayed are unitary evolution, SC regime (with $\gamma/\lambda = 0.1$) and WC regime (with $\gamma/\lambda = 2$). In SC regime, the path traced by $\ket{\Psi_+^{SC}(t)}$ differs from that traced by the unitarily evolved state $\ket{\psi(t)}$, yet, both states describe curves for all 7 periods. On the other hand, the eigenstate $\ket{\Psi_+^{WC}(t)}$ corresponding to the WC regime, while describing a curve which differs strongly from that traced in the unitary case, stops its motion in the Bloch sphere after a couple periods, consequently preventing a GP accumulation. For this reason, and as we have previously announced, we shall restrict our analysis to the SC regime.

\begin{figure}[ht]
\centering
\includegraphics[width=.9\columnwidth , trim={.75cm 0 .75cm 0}]{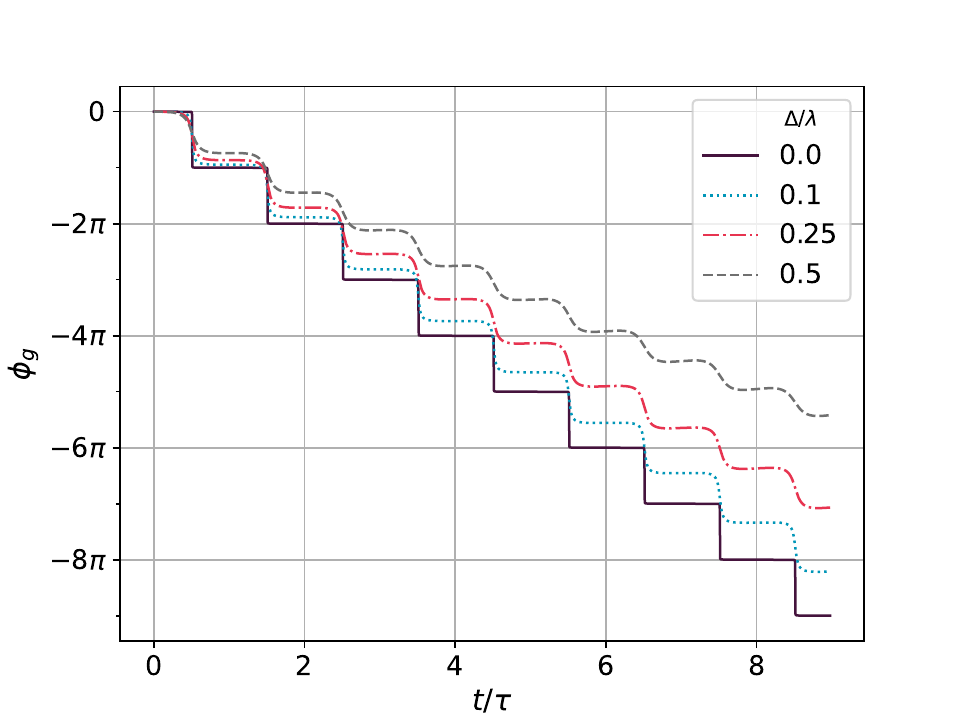}
\caption{Geometric phase $\phi_g (t)$ accumulated over time for different $\Delta/\lambda$ values, corresponding to different atom-mode normalized detuning. Parameter values are $\gamma/\lambda = 0.1, p/\lambda= 0.005$.}
\label{fig:FvsTDelta}
\end{figure}

It was shown in section \ref{sec:up} that in the unitary context, the one-parameter curve described in ray space as the system evolves depends considerably on the atom-mode detuning $\Delta$, and therefore, so does the GP. Dealing now with the case of an open system, Fig. \ref{fig:FvsTDelta} shows the phase accumulated over time for different $\Delta/\lambda$ values, exposing a manifest dependence of $\phi_g$ on $\Delta/\lambda$, as well as in the unitary case. In particular, it can be seen that as the value of $\Delta/\lambda$ is increased, the geometric phase accumulated over time decreases (in absolute value) and softens.

\begin{figure}[ht]
\centering
\includegraphics[width=.9\linewidth, trim = {1cm 0 1cm 0}]{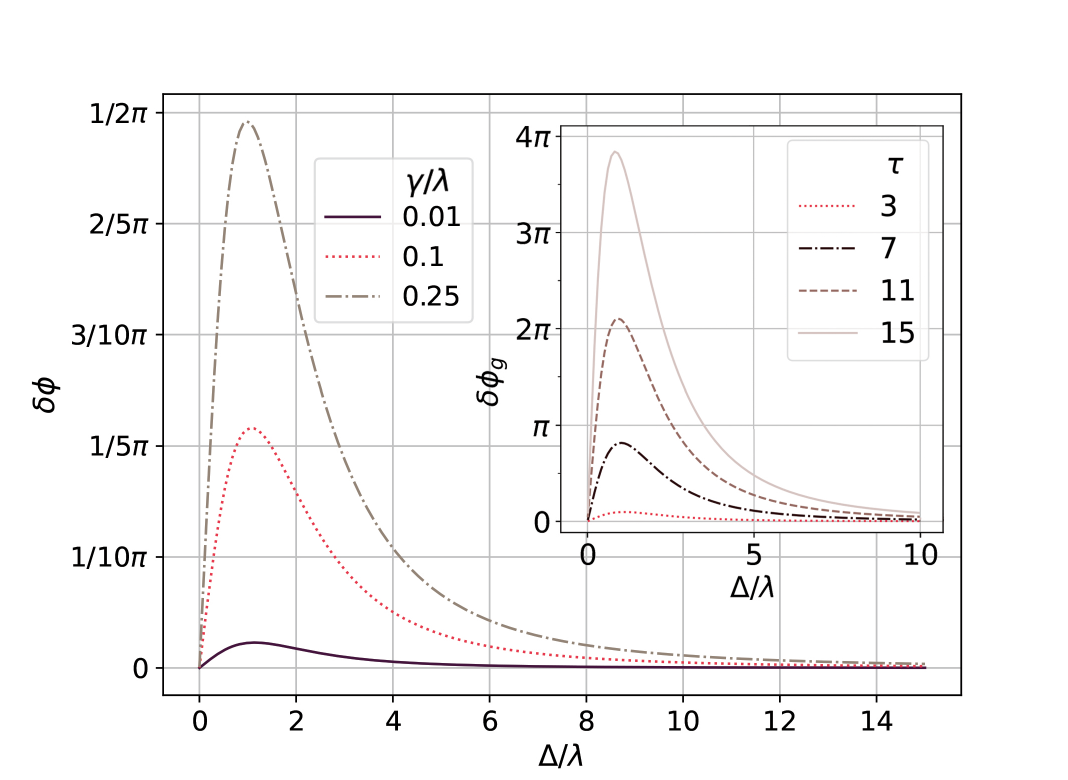}
\caption{Correction $\delta\phi = \phi_g-\phi_u$ introduced in the GP by the presence of the environment as a function of the atom-mode normalized detuning $\Delta/\lambda$, at a fixed time $t= 3\tau$ and for different values of $\gamma/\lambda$. The inset shows the same relation $\delta\phi(\Delta/\lambda)$ for different time intervals, at fixed $\gamma/\lambda = 0.1$. In both the main plot and the inset, the amplitude of the  continuous pump is $p/\lambda= 0.005$}
\label{fig:FvsDelta}
\end{figure}

A natural question arising in this context is weather the effects introduced in the GP by the non-unitary evolution, embodied in the correction $\delta\phi$, depend on the normalized detuning $\Delta/\lambda$ themselves or whether the whole dependence on $\phi_g$  is contained in its unitary component $\phi_u$. In order to address this matter we plot, in Fig. \ref{fig:FvsDelta}, the correction $\delta\phi$ to the GP accumulated over three periods ($\tau = 3$) as a function of $\Delta/\lambda$. It can be seen that $\delta\phi$ indeed depends on  $\Delta/\lambda$, and two aspects of this dependence are straightforward to notice. The first is that the correction to the GP vanishes in the resonant case, making the resonant GP $\phi_g^{\Delta =0}$ robust to the non-unitary effects of the environment in the SC regime. The other characteristic of the correction $\delta\phi$ which is visible in Fig. \ref{fig:FvsDelta}, is that $\delta\phi$ reaches a maximum for some value of $(\Delta/\lambda)$. The value $(\Delta/\lambda)_\text{max}$ at which this maximum occurs, has a weak but non-vanishing dependence not only on the values of $\gamma$ and $p$, but also in the time at which the phase is observed, as it can be seen in Fig. \ref{fig:FvsTDelta} and in the inset included in it.

Hence, Fig. \ref{fig:FvsDelta} 
has a two-fold interest since allows the identification of both features: (i) those conditions which increase the effects of the environment on the geometric phase, bringing closer the possibility of detection, and (ii) those that mitigate and thus, enable to neglect them, or even eliminate any effect. While all the considerations regarding the robustness of the GP for the resonant case $\delta\phi^{\Delta =0}=0$ will be postponed up to next section, we will give here a pictorial view supporting the existence of a maximum for $\delta\phi(\Delta/\lambda)$.

\begin{figure}[h]
\centering
\includegraphics[width=.85\linewidth, trim={.5cm 0 .5cm 0}]{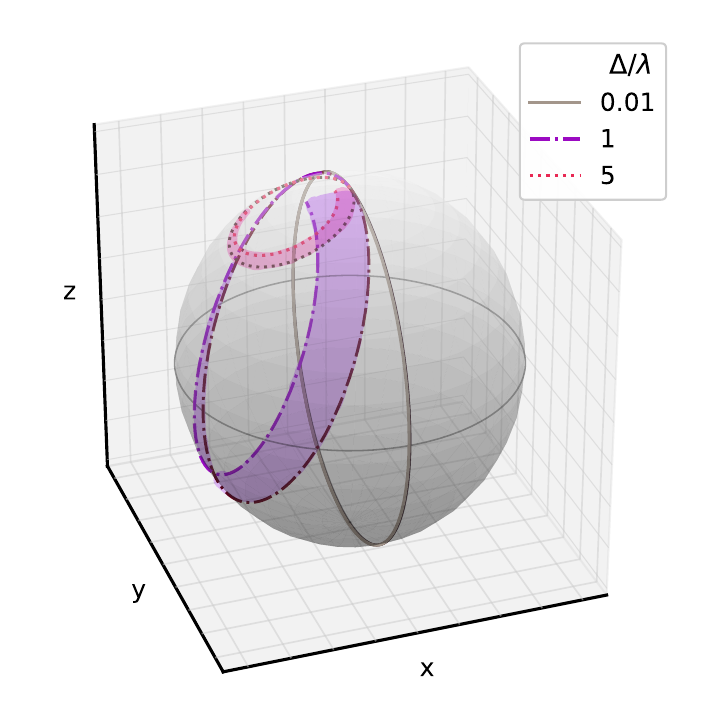}
\caption{Trajectories described in the Bloch sphere by the unitarily evolved state $\ket{\psi(t)}$ and the eigenstate $\ket{\Psi_+(t)}$ of the non-unitarily evolved density matrix $\rho(t)$ for $t\in [0, \tau]$ and for different values of the atom-mode detuning. Where the difference in the area subtended by each pair of curves, which is proportional to $\delta\phi$ is highlighted. Parameter values are $p/\lambda= 0.005$ and $\Delta/\lambda= 0.01, 1$ and $5$. A large value $\gamma/\lambda= 0.5$ was used to render the behaviour visible at plain sight.}
\label{fig:BlochMax}
\end{figure}

The greatest benefit of this view is attained when considering the widely extended and solidly shown geometrical interpretation of the unitary GP acquired by the state of a two-level system. That is, that $\phi_u$ is given by half the solid angle of the closed path traced in the Bloch sphere by a cyclic evolution, i.e., the area of the surface enclosed by the trajectory.
Recalling, on the other hand, the possibility of interpreting the non-unitary GP acquired by an initial state $\rho(0) = \ket{e, 0_c}\bra{e, 0_c}$ as the unitary GP acquired by the eigenstate $\ket{\Psi_+(t)}$ of $\rho(t)$, it is possible to observe the paths described by $\ket{\Psi_+(t)}$ and $\ket{\psi(t)}$ and compare the subtended areas, which are proportional to the GPs, for different $\Delta/\lambda$ values.
Fig. \ref{fig:BlochMax} shows both the unitary and the non-unitary trajectories described in a time $t=\tau$ for three $\Delta/\lambda$ values for which the environment influences the GP to considerably different extent. For $\Delta/\lambda = 0.1$ both trajectories enclose large but almost identical areas which subtraction, $\propto \delta\phi$, returns a small value. As $\Delta/\lambda$ grows and the unitary trajectories run on planes which are farther from the origin of the Bloch sphere, the difference between the area enclosed by the curves grows too, while the area enclosed by each one decreases. The greatest difference was always observed to fluctuate around $\Delta/\lambda =1$.  Eventually, these trajectories enclose smaller and smaller areas which difference decreases as well.

\subsection{Robustness of the resonant GP}\label{sec:resonant}
The robustness presented by the resonant GP $\phi_g^{\Delta =0}$ can be understood in geometric terms considering the evolution of the state $\rho(t)$ of the atom plus field as follows.
It is well known that it is possible to define geodesics on the ray space, i.e. curves of minimal length with respect to a canonical metric form \cite{mukunda93, book_phases}. For a two-level system, the geodesics of the Bloch sphere are its great-circles. Over geodesic curves, the expression in Eq. (\ref{fase:u}) vanishes for any pair of vectors for which it is well defined, that is, for any non-orthogonal $\ket{\psi(0)}$ and $\ket{\psi(t)}$ \cite{mukunda93, acotacion}. Thus, geodesics enable to understand the GP acquired in non-cyclic trajectory C in the ray space as that acquired in the associated close trajectory  $\tilde{\text{C}}=\text{C} \cup \text{C}^\text{geod}$, composed of C and the geodesic connecting its final and initial points. As the contribution of the evolution along the geodesic vanishes, the whole phase accumulated in describing this close curve be due to the evolution along C
\begin{equation}
     \phi_g[\tilde{\text{C}}] = \phi_g[\text{C}] + \phi_g[\text{C}^\text{geod}] = \phi_g[\text{C}].
\end{equation}

This idea is used to explain the $\pi$-jump exhibited by the GP when the trajectory described by the state of a two-level system on the Bloch sphere is half a great-circle. The sudden change in the geodesic for non-cyclic trajectories corresponding to almost-half a great-circle and little more than half a great-circle is the mathematical tool allowing to interpret this jumps, which have been both studied theoretically \cite{phasejumpth} and observed experimentally \cite{phasejumpexp}. Then, for this particular curves on the ray space, the GP will be null if the final state does not exceed the diametrically opposite point from the initial state and $\phi_g = \pi$ if it does.

On the other hand, the interpretation of Eq. (\ref{fase:iniu}) as the unitary GP acquired by the eigenvector $\ket{\Psi_+(t)}$ of the system's state $\rho(t)$ demands the consideration of the curve described by $\ket{\Psi_+(t)}$. It can be seen that, in each period $\tau$ of time, $\ket{\Psi_+(t)}$ moves over the same great-circle as $\ket{\Psi(t)}$ but it does so in an open trajectory, as it does not manage to reach the initial point. This is, however, of no importance for the computation of the GP, which takes the value of $\pi$ as long as the $\ket{\Psi_+(t)}$ manages to cover half the great-circle. 
Thus, as long as the dissipative effects are not strong enough to prevent the eigenket $\ket{\Psi_+^{\Delta = 0}(t)}$ to exceed the opposite pole for $t = \tau$, the GP will not be affected at all by the environment. The resonant case is consequently an ideal scene for those experimental tests which require to neglect dissipative effects.

Another way to understand the robustness of the resonant GP $\phi_g^{\Delta =0}$ relays on the conjunction of two aspects of the dynamics which have already been mentioned separately, namely, the invariance of the GP under the action of static (non)unitary operators on the ray space, and the fact that the equatorial trajectory corresponds to an evolution restricted to maximally entangled states of a pair of qubits. Hence, as already discussed in Sec.\ref{sec:up}, the phase acquired in the resonant $\Delta = 0$ scenario matches, up to a minus sign, with that acquired by a two-qubit system, when its evolution is restricted to maximally entangles states. For these particular evolutions, it has been consistently shown thae robustness of the GP \cite{OXMANposta, Oxman_fractional}.

\section{Conclusions}\label{sec:conclu}
We present in this article a new way of examining the non-unitary effects which are present in dissipative JC model: the evaluation of the GPs acquired by a state of the atom-cavity system. A detailed study of the GP is carried on, and the frameworks over which comparison with the unitary JC model is plausible are discussed. After over-viewing the usual approach to GPs in JC model, which consists on the computation of the Berry phase acquired by the eigenstates of JC Hamiltonian when a field phase-shift operator acts adiabatically on the system, we have introduced Mukunda and Simon's kinematic approach to GPs. This approach, in which the GP depends exclusively on the trajectory described in the ray space, allows for the study of GPs emerging in the dynamical evolution generated by JC hamiltonian, in which no Berry phase is accumulated. It also enables a simple explanation for the coincidence up to a sign we find between the GP accumulated by the $\ket{e, 0_{\text{c}}}$ in the evolution given by $\hat{H}_{\text{JC}}$ and the usual result for the Berry phase acquired by eigenstate $\ket{\Phi_0^+}$, as the trajectories described in the Bloch sphere during these revolutions are related by ray space isometries.
After analyzing the case in which the system evolves isolated from its environment, we have dealt with the problem of the description of GPs for a dissipative JC model. We solved the phenomenological master equation, which takes into account both the leakage of photons through the cavity mirrors and the pumping of the two-level system, and numerically obtained the state of the system at any time. By the inspection of the atom-mode state and the evolution of its eigenstate $\ket{\Psi_+(t)}$ on the Bloch sphere, we have determined that the WC regime, characterized by the relation $\gamma/\lambda > 1$ between the dissipation rate of the system $\gamma$ and the interaction constant $\lambda$, is not suitable for any study of GPs, as thus we showed that the system losses coherence too fast for any phase to be accumulated. 
Finally, we focused on the SC regime, where $\gamma/\lambda < 1$ and the coherence is maintained for long enough periods.
In that regime, we found the correction $\delta\phi$ introduced in the GP $\phi_g = \phi_u + \delta\phi$ by the occurrence of dissipate effects increases, as expected, as $\gamma/\lambda$ does. But we also found that, at a fixed time and for each fixed value of the parameters $\gamma/\lambda$ and $p/\lambda$, there is a value $\Delta/\lambda$ of the atom-mode detuning that maximizes this correction, thus setting ideal conditions for detection. The existence of a maximum in the GP correction $\delta\phi$ as a function of $\Delta/\lambda$ was also supported on the grounds of geometrical considerations.
Finally, we studied the robustness of the resonant case $\Delta=0$, meaning  that the phase acquired by the state of the open quantum system identically coincides with that acquired in the unitary case $\phi_g = \phi_u$ independently of the relation $\gamma/\lambda$. We have demonstrated this feature with certainty. 

It is worth mention that the scheme under study arises not only in the context of realistic cavity electrodynamics but also in scenarios such as circuit QED, in which JC model is used to explain the coherent coupling of superconducting qubits to microwave photons \cite{circuitQED_review}. As any open quantum system, the coupling of superconducting circuits to their environment, which is necessary for coherent control and measurements in circuit QED, invariably leads to decoherence. Therefore circuit QED appears as a natural architecture to measure the corrections in the GP in disipative JC models.

\section*{Acknowledgements}
The authors would like to thank L.E. Oxman for insightful discussions. L.V. acknowledges ICTP Sandwich Training
Educational Programme (STEP). 
This work was supported by ANPCyT, CONICET, and Universidad de Buenos Aires; Argentina.

\bibliography{0casimir.bib,0cuantica.bib,0teoria.bib}

\begin{thebibliography}{58}
\bibitem [{aco()}]{acotacion}%
  \BibitemOpen
  \href@noop {} {\ }\bibinfo {note} {A clarification that may be made at this
  point is that both the first term in the r.h.s. of Eq. (\ref{fase:u}) and
  hence $\phi_g[\text{C}]$ itself are undefined in case $\ket{\psi({0})}$ and
  $\ket{\psi{(t)}}$ are orthogonal to each other.}\BibitemShut {Stop}%
\end{thebibliography}

\end{document}